\documentclass[preprintnumbers,amsmath,amssymb,floatfix,11pt,prd,onecolumn,
superscriptaddress,nofootinbib]{revtex4}
\usepackage{graphicx}
\usepackage{epsfig}
\usepackage{bm}
\usepackage{amsfonts}
\begin{document}

\title{Conformal Invariant  Teleparallel Cosmology}

\author{\textbf{Davood Momeni}}
\author{\textbf{Ratbay Myrzakulov}}
\affiliation{Eurasian International Center for Theoretical Physics, Eurasian National University, Astana 010008, Kazakhstan}

\begin{abstract}
\textbf{Abstract:}We study teleparallel gravitational theories with are invariant under the conformal transformations. Wide family of the gravitational Lagrangians that are invariant under conformal transformations have investigated. Cosmological solutions inspired by the observational data for a flat Universe in the vacuum has been found. To add matter fields to the cosmological models by preserving the conformal symmetry there are significant limitations. Friedmann-Robertson-Walker (FRW)  equations based on the effective terms of energy density and pressure have been rewritting and continuty equation for the effective quantities are derived. In the vacuum and in the absence of any matter field,the FRW equation has an exact solution for Hubble parameter which is consistent with the cosmological data,specially given analytical solution is in good agreement with $\Lambda$CDM model in the present time. Then the scalar field-Torsion models in the cosmological FRW background investigated. Massless scalar field equations are very complex with an exact analytical solution in special limits. We have shown that the Lagrangian scalar field with self interaction $V(\phi)=\frac{1}{4!}\mu\phi^4$ can be written that the conformal symmetry is preserved.

\textbf{Keywords:}Tetrad theory of gravity; Torsion; Conformal invariance.

\end{abstract}
\maketitle

\section{Introduction}
Gravity is a gauge theory that any formulation of gravity should remains invariant under a special class of gauge transformations, from this perspective gravity is similar to electrodynamics. Now, complete and accurate description of gravity that able to describe the Universe on large scales does not exist. Also there is no realiable explanation for the fundamental forces of the nature. Along with providing  general relativity (GR) another model for gravity introduced by Einstein in which the gravity was not curvature effect of the spacetime but the torsion of the spacetime explained gravity. In this formulation we need to rewrite gravity by tetrads language. With this new formulation many similiarities appear between gravity and gauge theory of electrodynamics. Equivalence between GR and gravity description in tetrads language called teleparallel
\cite{Maluf2,Hehl,Poplawski}.  The idea of teleparallelism in the last century has been studied from various aspects. 
\par
Conformal invariance is one of the most important concepts in  modern physics. This concept was originally introduced to explain 
high energy physics  and later went to work in inflationary cosmology. Conformal symmetry is a local transformation that acts on a metric of spacetime and if the gravitational theory is not Lorentz invariant,is significant. Conformal symmetry in gauge theory of gravity has central role \cite{tHooft1,tHooft2}. 
\par
Conformal symmetry can be considered as an internal symmetry of spacetime. Recently it has been shown that conformal symmetry can be interpreted as an alternative model for gravity that gives a better description for inflation \cite{Chamseddine:2013kea,Chamseddine:2014vna}.
\par Gravitational theories including higher order terms of curvature tensor with respect to the conformal symmetry have been studied many times. As a first example Weyl's theory with conformal symmetry in Ref.\cite{Mannheim1,Mannheim2,Moon2} have been investigated.
\par
It is a question that can we write teleparallel theory for gravity which is conformal invariant?. This means that there is at least one gravitational Lagrangian that can preserve conformal symmetry. In the framework of teleparallel gravity, recently has been introduced a mechanism which can be written teleparallel gravity as a conformal invariance
 \cite{Maluf1,Maluf:2012yn}. Specially it is shown how speciefic combinations of torsion tensor can be identified as basic blocks to construct teleparallel with conformal symmetry. Our main objective in this paper is that cosmological aspects of such gravitational models should be carefully checked.\par
Notation: Greek indices $\mu, \nu, ...$ ,spinor indices $a, b, ...$
run from 0 to 3,
$\mu=0,i,\;\;a=(0),(i)$. We use the covariant representation of the tetrad field  $e^a\,_\mu$. The fundamental  tensor 
is defined by $T_{a\mu\nu}=\partial_\mu e_{a\nu}-\partial_\nu e_{a\mu}$., $e=\det(e^a\,_\mu)$.

The torsion tensor is defined by 
$T^a\,_{\mu\nu}$. We mention here that the torsion tensor is related to the  Weitzenb\"ock  connection 
$\Gamma^\lambda_{\mu\nu}=e^{a\lambda}\partial_\mu e_{a\nu}$. The spacetime manifold $(\mathcal{M},g)$ is 
the Weitzenb\"ock spacetime. By the defintion, we set to the zero the curvature (constructed from the  Weitzenb\"ock connection) . We work in the framework of the  Riemannian and Weitzenb\"ock geometries. 

Our plan in this paper is as follows. In Sec. 2 we construct the teleparallel gravity in favor of the conformal invariance. In Sec. 3 we study the FRW cosmology for a simple class of the models. In Sec. 4 we study the scalar-torsion models. We conclude in Sec. 5 .

\section{Construction of the conformal invariance  Teleparallel Lagrangian}
In Ref.\cite{Maluf1,Maluf:2012yn} proposed teleparallel gravity models in which Lagrangian based on the fundamental properties of conformal transformations. Our aim in this section is a mechanism for writting such Lagrangians.
\par
In gravitational physics,conformal transformation of a Lorentzian metric  $(\mathcal{M},g)$ is a local transformation that acts on the metric
$g_{\mu\nu}$ and converts it to the new metric $\tilde{g}_{\mu\nu}=\Omega(x)^2g_{\mu\nu}$, 
in such transformations $\Omega(x)$ is a piecewise continous and non singular function entirely spacetime. It is assumed that the second order derivatives of $\Omega(x)$ is well defined. In this framework, we are required to provide a set of orthogonal basis vectors $e_{a\mu}$ so that the metric is written as  :
\begin{equation}
g_{\mu\nu}=e_{a\mu}e_{b\nu}\eta^{ab}
\end{equation}
By applying conformal transformations into metric the transformation spreads to each of the following basis:
\begin{equation}
\tilde{e}_{a\mu}=\Omega(x)\,e_{a\mu}\,, \ \ \ \ \ \ \ \ \ \ \
\tilde{e}^{a\mu}=\Omega(x)^{-1}\,e^{a\mu}\,.
\end{equation}
With above transformation,we are able to build major blocks of   Weitzenb\"ock  spacetime with torsion, the first quantity is:
\begin{eqnarray}
T_{abc}=e_b\,^\mu e_c\,^\nu (\partial_\mu e_{a\nu}-\partial_\nu e_{a\mu}).
\end{eqnarray}
Transformation property of basis $\tilde{e}_{a\mu}$ leads to the following transformations for $T_{abc}$:
\begin{eqnarray}
\tilde{T}_{abc}&=&
\Omega^{-1}(x)(T_{abc} + \eta_{ac}\,e_b\,^\mu \partial_\mu \log\Omega
-\eta_{ab}\,e_c\,^\mu \partial_\mu \log\Omega)\,, \nonumber \\
\tilde{T}^{abc}&=&
\Omega^{-1}(x)(T^{abc} + \eta^{ac}\,e^{b\mu} \partial_\mu \log\Omega
-\eta^{ab}\,e^{c\mu} \partial_\mu \log\Omega)\,. 
\label{2}
\end{eqnarray}
The rank three tensor can be defined the following vector quantity:
\begin{eqnarray}
T_a=T^b\,_{ba}.
\end{eqnarray} 
Under transformations,the vector will change as follows:
\begin{eqnarray}
\tilde{T}_a&=&\Omega^{-1}(x)(T_a-3\,e_a\,^\mu \partial_\mu \log\Omega)\,, 
\nonumber \\
\tilde{T}^a&=&\Omega^{-1}(x)(T^a-3\,e^{a\mu} \partial_\mu \log\Omega)\,.
\label{3}
\end{eqnarray}
From above vector quantity we can define a vector in the spacetime frame:
\begin{eqnarray}
T_{\mu}=e_{a\mu}T^{a}.
\end{eqnarray}
Using the transformation properties of the basis we have: 
\begin{equation}
\tilde{T}_\mu= T_\mu -3\partial_\mu \log\Omega
\end{equation}
The expression   $T_\mu$ is not only possible to construct a vector in the spacetime frame and other components can be tested. However,untill we have reviewed ,$T_\mu$ is the simplest possible.\par
Now we able to introduce Lagrangian term which remains invariant under conformal transformations. For a better understanding ,we consider the following series of scalar quantities under the conformal transformation:

\begin{eqnarray}
\tilde{T}^{abc}\tilde{T}_{abc}&=& \Omega^{-2}(x)(T^{abc}T_{abc}-
4T^\mu \partial_\mu\log \Omega(x)+6g^{\mu\nu}\partial_\mu\log\Omega(x)\partial_\nu \log\Omega(x))\,,
\nonumber \\
\tilde{T}^{abc}\tilde{T}_{bac}&=& \Omega^{-2}(x)(T^{abc}T_{bac}-
2T^\mu \partial_\mu\log\Omega(x)+3g^{\mu\nu}\partial_\mu\log\Omega(x)\partial_\nu \log\Omega(x))\,,
\nonumber \\
\tilde{T}^a\tilde{T}_a&=&\Omega^{-2}(x)(T^a\,T_a -6T^\mu \partial_\mu \log\Omega(x)
+9g^{\mu\nu}\partial_\mu\log\Omega(x)\partial_\nu \log\Omega(x))\,.
\end{eqnarray}
The simplest combination of the above terms as follows is essential for making invariant Lagrangian and plays the key role in our theory:
\begin{equation}
\tau={1\over 4}T^{abc}T_{abc}+{1\over 2}T^{abc}T_{bac}-{1\over 3}T^aT_a\label{tau}\,
\end{equation}
The expression (\ref{tau}) converts into
$\tilde{\tau}=\Omega^{-2}(x)\tau$ and the most simple proposal is the following teleparallel Lagrangian:
\begin{eqnarray}
\mathcal{L}=e\tau^2\label{L1}.
\end{eqnarray}
It is straighforward ro show that expression (\ref{L1}) is invariant :
\begin{eqnarray}
\tilde{\mathcal{L}}=\tilde{e}\tilde{\tau}^2=(\Omega^{4}(x)e)(\Omega^{-2}(x)\tau)^2=\mathcal{L}.
\end{eqnarray}
Another Lagrangian proposed by Maluf,which includes auxiliary scalar  field  $\phi$ which changes under conformal transformation as follows:
\begin{equation}
\tilde{\phi}=\Omega^{-1}(x) \phi.
\end{equation}
Lagrangian considering the scalar field (one dgree of freedom) is as follows:
\begin{equation}
{\cal L} =  e\biggl[
-\phi^2\tau
+k'g^{\mu\nu}D_\mu \phi D_\nu \phi\biggr]\label{L2}.
\end{equation}
 $k'$ in the (\ref{L2}) is a free coupling constant and the derivative for the scalar field is defined as follows:

\begin{equation}
D_\mu \phi=\biggl(\partial_\mu -{1\over 3} T_\mu \biggr)\phi\,.
\label{9}
\end{equation}
It is not hard to show that  $\tilde{D}_\mu \tilde{\phi}=\Omega^{-1}\,D_\mu \psi$. Then (\ref{L2}) is invariant under conformal transformations.\par
The Lagrangians (\ref{L1}),(\ref{L2}) are not only examples for conformal invariances,for example following lagrangian is invariant :
\begin{equation}
{\cal L}(e_{a\mu})= e  L_1 L_2 \,,
\label{11}
\end{equation}
In which:
\begin{equation}
L_1=A\, T^{abc}T_{abc}+ B\, T^{abc}T_{bac}+ C\, T^aT_a\,,
\label{12}
\end{equation}
\begin{equation}
L_2=A'\, T^{abc}T_{abc}+ B'\,T^{abc}T_{bac}+ C'\, T^aT_a\,.
\label{13}
\end{equation}
The only constraints are that coefficents $\{A,B,C,A',B',C'\}$ must be satisfy in the following equations:
\begin{equation}
2A+B+3C=0\,, \ \ \ \ \ \ \ \ \ \ 2A' +B' +3C'=0\,.
\label{14}
\end{equation}
Our main goal in this paper is to finding cosmological solutions for  (\ref{L1},\ref{L2}).
\section{Cosmological solutions of FRW for $
\mathcal{L}=e\tau^2$}
In this section we will investigate the equations of motion for model (\ref{L1}). Inspired by the cosmological data,the simplest model of Universe is a flat and homogenous and isotropic model  FRW.  Simple and suitable tetrads basis for metric is:
\begin{eqnarray}
&&e^{a}_{\mu}=diag(1,a(t),a(t),a(t)).
\end{eqnarray}
 non-zero components of  fundamental tensor are :
\begin{eqnarray}
&&T_{i\mu\nu}=a\dot{a}\Big(\delta_{\mu 0}\delta_{\nu 1}-\delta_{\nu 0}\delta_{\mu 1}\Big),\ \ i=1,2,3.
\end{eqnarray}
 components can be expressed as follows:
\begin{eqnarray}
&&T_{101}=-T_{110}=a\dot{a},\\
&&T_{202}=-T_{220}=a\dot{a}\\
&&T_{303}=-T_{330}=a\dot{a}.
\end{eqnarray}
components of the vector $T_{\nu}$ can be immediately written as:
\begin{eqnarray}
&&T_{\nu}=-3H\delta_{\nu 0},\ \ H=\frac{\dot{a}}{a}.
\end{eqnarray}
give the  $T_{a}=e_{a}^{\mu}T_{\mu}$ we can write:
\begin{eqnarray}
&&T^aT_a=-9H^2.
\end{eqnarray}
We should compute  components of $T_{abc}$ for the final form of $\tau$:
\begin{eqnarray}
&&T_{1 bc}=a^2\dot{a}\Big(\delta_{b 0}
\delta_{c 1}-\delta_{b 1}\delta_{c 0}\Big),\\
&&T_{2 bc}=a^2\dot{a}\Big(\delta_{b 0}
\delta_{c 2}-\delta_{b 2}\delta_{c 0}\Big),\\
&&T_{3 bc}=a^2\dot{a}\Big(\delta_{b 0}
\delta_{c 3}-\delta_{b 3}\delta_{c 0}\Big).
\end{eqnarray}
Using these components we can list elements of $T^{abc}$:
\begin{eqnarray}
&&T^{abc}=a^2\dot{a}\delta^{a i}\Big(\delta_{b i}
\delta_{c 0}-\delta_{b 0}\delta_{c i}\Big),\ \ i=1,2,3.
\end{eqnarray}
Finally,$T^{abc}T_{abc}$ can be written as follows:
\begin{eqnarray}
&&T^{abc}T_{abc}=-6(a^2\dot{a})^2.
\end{eqnarray}
Using the symmetry properties of $T_{abc}$ we can calculate the following: 
\begin{eqnarray}
&&T^{abc}T_{bac}=-6(a^2\dot{a})^2.
\end{eqnarray}
and finally obtained :
\begin{eqnarray}
&&\tau=H^2(3-\frac{9}{2}a^6).
\end{eqnarray}
point like form of (\ref{L1}) in the FRW Universe have  simple form as follows:
\begin{eqnarray}
&&\mathcal{L}=e\tau^2=f(a)\dot{a}^4,\\
&&f=f(a)=\frac{(3-\frac{9}{2}a^6)^2}{a}.
\end{eqnarray}
Here's a type about the matter in the FRW equations. First , we note that the perfect fluid with barotropic equation of state (EoS) $p=w\rho$,breaks the conformal symmetry. If you want to add matter fields to the (\ref{L1}), should be added a Lagrangian form (\ref{L1}) which is respect to conformal symmetry. It means the Lagrangian of matter fields must be invariant under the transformations: 
\begin{eqnarray}
\tilde{\mathcal{L}}_m=\mathcal{L}_m.
\end{eqnarray}
As the simplest example,radiation is a simple and handy but it is not only example of such  matters. Infact,to make the FRW Universe,can build a simple form of Lagrangian of matter fields. Such a simple form is as follows:
\begin{eqnarray}
\mathcal{L}_m=\rho_{m0}  a(t)^4.
\end{eqnarray}
Above Lagrangian is invariant because as we know 
 $$
e\to \tilde{e}=\Omega(x)^4 e,\ \ a(t)\to\tilde{a}(t)=\Omega(x)^{-1}a(t)
$$
As a result:
\begin{eqnarray}
\tilde{\mathcal{L}}_m=\rho_{m0} (\tilde{e} \tilde{a(t)}^4)=\rho_{m0} \Omega(x)^4 e(\Omega(x)^{-1}a(t))^4=\mathcal{L}_m.
\end{eqnarray}
\par
If we want solve FRW equations for (\ref{L1}),the solution is complex. To avoid mathematical complexity,we confine ourselves to the 
$\mathcal{L}_m=0$.
 \par
and we find accurate solutions for the equations. The modified FRW equations can be written in terms of effective quantities:
\begin{eqnarray}
&&3H^2=\rho_{eff},\label{frw1}\\
&&2\dot{H}+3H^2=-p_{eff}\label{frw2}.
\end{eqnarray}
in which:
\begin{eqnarray}
&&\rho_{eff}=-\int{dt\Big[H^3(2+\frac{fa}{2f'})\Big]},\label{rho-eff}\\
&&p_{eff}=-H^2(1-\frac{fa}{2f'})\label{p-eff}.
\end{eqnarray}
 equation (\ref{frw1}) and by using the (\ref{frw2}) 
\begin{eqnarray}
&&\dot{\rho}_{eff}+3H(p_{eff}+\rho_{eff})=0
\end{eqnarray}
\par
Now we annalyze the continuty equation of this particular model.  Continuty equation for a perfect fluid in any model of cosmology which has FRW equations (\ref{frw1}),(\ref{frw2}) is as follows:
\begin{eqnarray}
&&\dot{\rho}_{eff}+3H(\rho_{eff}+p_{eff})=0.
\end{eqnarray}
In case (\ref{L1}) regardless of the gravity model is written based on torsion,the form of continuty equation is written without changing. Therefore, with effective quatities $(\rho_{eff},p_{eff})$ the energy conservation equation remains unchanged. To find the exact solution of FRW in the vacuum,we rewrite the second FRW as follows:
\begin{eqnarray}
&&\ddot{a}+\frac{1}{4}\frac{f}{f'}\dot{a}^2=0\label{eom1}.
\end{eqnarray}
By change of variable $X=Ha$, equationo (\ref{eom1}) become as follows:
\begin{eqnarray}
&&\frac{dX}{da}+\frac{1}{4}\frac{f'(a)}{f(a)}X=0\label{Xeq}.
\end{eqnarray}
And the exact solutions of (\ref{Xeq}) is :
\begin{eqnarray}
&&X(a)=\sqrt[4]{\frac{c_1}{|f(a)|}}.
\end{eqnarray}
It is convenient to rewrite the solution in terms of  the redshift $1+z=\frac{1}{a}$,
\begin{eqnarray}
&&H(z)=\sqrt[4]{c_1}\frac{(1+z)^{3/4}}{\sqrt{|3-\frac{9}{2(1+z)^6}|}}
\end{eqnarray}
The cosmological solution of model (\ref{L1}) is vacuum. Thus,the compatiability of our model with realistic models we compare the solution $H(z)$ with 
$\Lambda$CDM. 
\begin{eqnarray}
&&H^2(z)|_{z=0}=\frac{\sqrt{c_1}}{\frac{3}{2}}=H_0^2,\ \ c_1=(1.5 H_0^2)^2.
\end{eqnarray}
Our model at $z=0$ gives:

\begin{eqnarray}
&&\frac{H^2(z)}{H_0^2}=\frac{1.5(1+z)^{3/2}}{|3-\frac{9}{2(1+z)^6}|}
\end{eqnarray}

In the vicinity of $z=0$:
\begin{eqnarray}
&&\frac{H^2(z)}{H_0^2}|_{z=0}=1+(0.195) z+(0.288)z^2+\mathcal{O}(z^3)
\end{eqnarray}

$\Lambda$CDM is:

\begin{eqnarray}
&&3H^2(z)=\rho_{m0}(1+z)^3+\Lambda.
\end{eqnarray}
Dimensionless the Hubble parameter is to devide both sides of equation on $H_0^2$ and :
\begin{eqnarray}
\Omega_{m0}=\frac{\rho_{m0}}{3H_0^2},\ \ \Omega_{\Lambda 0}=\frac{\Lambda}{3H_0^2}.
\end{eqnarray}
So:
\begin{eqnarray}
&&\frac{H^2(z)}{H_0^2}=\Omega_{m0}(1+z)^3+\Omega_{\Lambda 0}.
\end{eqnarray}
That:
\begin{eqnarray}
\Omega_{m0}\sim0.75,\ \ \Omega_{\Lambda 0}\sim0.25.
\end{eqnarray}
In the vicinity of $z=0$,$\Lambda$CDM gives the following expression:
\begin{eqnarray}
&&\frac{H^2(z)}{H_0^2}|_{z=0}=1+(0.375)z+(0.305)z^2+\mathcal{O}(z^3).
\end{eqnarray}
Thus in the prsent time our model is in good agreement with $\Lambda$CDM and the final graph shows a good match. For better comparison with the $SneIa+BAO+CMB$ data ,the $\mu(z)$ have been plotted in terms of the theoretical result of $H(z)$. 
\begin{figure*}[thbp]
\includegraphics[width=6.5cm]{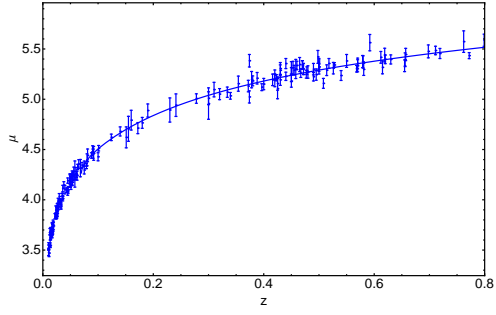}
\caption{Observational data  $SneIa+BAO+CMB$ for distance modulus $\mu(z)$ versus our theoretical result (\ref{Hz}) .}
\end{figure*}\par

\section{Cosmology of scalar-torsion fields}
The second example is a scalar-torsion which is given by
 (\ref{L2}). Here we are reminded that the (\ref{L2}) remains invariant by $V(\phi)=\frac{1}{4!}\mu\phi^4$,however obtaining accurate solution for such model by $V(\phi)$ is a difficult task and so we limited talk to a scalar field without potential. Lagrangian point like FRW is as follows:
\begin{eqnarray}
&&\mathcal{L}=a^3\Big[\phi^2h(a)H^2+k'(\dot{\phi}+H\phi)^2\Big],\\
&&h=h(a)=3-\frac{9}{2}a^6.
\end{eqnarray}
Equations of motion are modified Klein-Gordon by covariant derivative and both equations are FRW:
\begin{eqnarray}
&&\ddot{\phi}+3H\dot{\phi}+\phi\Big(\frac{\ddot{a}}{a}+\frac{H^2}{k'}(k'+h(a))\Big)=0,\\
&&\frac{\ddot{a}}{a}(h(a)+k')=-\frac{H^2}{2}\Big(k'+h(a)+ah'(a)\Big)-\frac{2H\dot{\phi}}{\phi}(k'+h(a))-k'\frac{\ddot{\phi}}{\phi}+\frac{k'}{2}(\frac{\dot{\phi}}{\phi})^2.
\end{eqnarray}
The first FRW equation is as follows:
\begin{eqnarray}
&&3H^2=\rho_{eff},\\&&
\rho_{eff}=\int{\Big[3H^3\frac{k'-h-ah'}{h}+\frac{6H^2\dot{\phi}}{h\phi}(k'-2h)+\frac{3k'H^2\dot{\phi}^2}{h\phi^2}\Big]dt}  .
\end{eqnarray}
To find an analytical solution of a analytical function,we define an analytical assisted auxiliarly function as follows:
 $x=a\phi$:
\begin{eqnarray}
&&\mathcal{L}=x^2ah(a)\dot{a}^2+k'a\dot{x}^2.
\end{eqnarray}
For further simplifications ,effective tortoise coordinate is defined as follows:
\begin{eqnarray}
&&b:=\int{a\sqrt{h(a)}da},\ \ z:=\sqrt{k'}\log x.
\end{eqnarray}
And the simplified configuration Lagrangian is:
\begin{eqnarray}
&&\mathcal{L}=a(b)e^{2z/\sqrt{k'}}(\dot{b}^2+\dot{z}^2).
\end{eqnarray}
Euler-Lagrange equations  for $\{b,z\}$ can be written as follows:
\begin{eqnarray}
&&\ddot{b}+\frac{2\dot{b}\dot{z}}{\sqrt{k'}}-\frac{\dot{z}^2}{a\sqrt{h}}=0,\\
&&\ddot{z}+\frac{\dot{b}\dot{z}}{a\sqrt{h}}-\frac{2\dot{b}^2}{\sqrt{k'}}=0.
\end{eqnarray}
\par
Above equations are severaly dissipative and fully non linear and the only possible solution for the equations is numerically. An exact solution for the case slowly varying $z$ field ($\dot{b}\dot{z}\gg\ddot{z}$) is obtained by solving the following equation and analytical methods in the inflationary models:
\begin{eqnarray}
&&\ddot{b}+\Omega\sqrt{b}\dot{b}^2\sim 0,\ \ \Omega=\frac{2}{\sqrt{k'k''}}-\frac{1}{\sqrt{2\sqrt{3}}k''},\ \ k''=\frac{k'}{8}.
\end{eqnarray}
which has non specific answer as follows:
\begin{eqnarray}
&&-b ( t) \Gamma( 2/3) {12}^{2/3}+b( t
) \Gamma ( 2/3,-2/3\Omega b( t ) 
^{3/2}) {12}^{2/3}+6C_1t\sqrt [3]{b
 ( t) ^{3}{\Omega}^{2}}+6C_2\sqrt [3]{
  b( t) ^{3}{\Omega}^{2}}=0,\\&&
z(t)=\frac{2}{3\sqrt{k''}}b(t)^{3/2}.
\end{eqnarray}
we remind that there is no simple way to find general solutions of $\{z,b\}$.

\section{summary}
Conformal transformations which are make conformal symmetry are a large family of physical models. Conformal symmetry means a rescaling metric and tetrads basis of spacetime. Various quantities are trap in two different forms in the teleparallel gravity. In this paper we have studied the conformal symmetry of teleparallel model inspired by recent studies regarding the conformal symmetry. We extend one of the torsion scalar modes to the potential case and the potential equation is $V(\phi)=\frac{1}{4!}\mu\phi^4$. Cosmological solutions of the families of teleparallel were studied in FRW equations. The metter can not be added to gravity Lagrangian in any form because the symmetry is broken. FRW equations have accurate analytical solutions and appllicable to the cosmilogical data and $Lambda$CDM model.  Analytical solution is obtained is scalar model and certain limits. Here we recall recently inspired by $f(R)$
\cite{f(R)}have been considered some models of gravity with torsion which are known $f(T)$ models \cite{f(T)},\cite{Jamil:2012fs} . It has been shown that  \cite{Yang} $f(T)$ is not conformal invariant but a simple generalization of $f(T)$ is presented which is conformal symmetric \cite{PLB}. Our research opens a window for constructing models of $f(T)$ which are invariant conformal transformations.

\section{ACKNOWLEDGEMENTS}
We are indebted   J.~W.~Maluf for many useful communications, we would like to thank  the anonymous reviewer for enlightening comments related to this work.


\begin{thebibliography}{99}

\bibitem{Maluf2}
J. W. Maluf, Gen. Rel. Grav. {\bf 19}, 57 (1987). 
\bibitem{Hehl}
F. W. Hehl, J. D. McCrea, E. W. Mielke and Y. Ne'eman, Phys. Rep. {\bf 258},
1 (1995).

\bibitem{Poplawski}
N. J. Poplawski, Mod. Phys. Lett A {\bf 22}, 2701 (2007); {\bf 24}, 431
(2009).

\bibitem{tHooft1}
G. 't Hooft, ``Probing the small distance structure of canonical quantum
gravity using the conformal group" [arXiv:1009.0669]

\bibitem{tHooft2}
G. 't Hooft, ``The conformal constraint in canonical quantum gravity",
[arXiv:1011.0061]; Found. Phys. {\bf 41}, 1829 (2011) [arXiv:1104.4543].


\bibitem{Chamseddine:2013kea} 
  A.~H.~Chamseddine and V.~Mukhanov,
  JHEP {\bf 1311}, 135 (2013)
  [arXiv:1308.5410 [astro-ph.CO]].

\bibitem{Chamseddine:2014vna} 
  A.~H.~Chamseddine, V.~Mukhanov and A.~Vikman,
  arXiv:1403.3961 [astro-ph.CO].


\bibitem{Mannheim1}
P. D. Mannheim, Prog. Part. Nucl. Phys. {\bf 56}, 340 (2006).

\bibitem{Mannheim2}
P. D. Mannheim, ``Making the case for conformal gravity" [arXiv:1101.2186].

\bibitem{Moon2}
T. Moon, P. Oh and J. Sohn, JCAP 1011, 005 (2010) [arXiv:1002.2549].

\bibitem{Maluf1}
J. W. Maluf and F. F. Faria, Phys. Rev. D {\bf 85}, 027502 (2012).

\bibitem{Maluf:2012yn} 
  J.~W.~Maluf and F.~F.~Faria,
  Annalen Phys.\  {\bf 524}, 366 (2012)
  [arXiv:1203.0040 [gr-qc]]..


\bibitem{f(R)}
H. A. Buchdahl, Mon. Not. Roy. Astron. Soc., 150, 1 (1970).
\bibitem{f(T)}
G.R. Bengochea , R. Ferraro 2009, Phys. Rev. D 79, 124019.
\bibitem{Jamil:2012fs} 
  M.~Jamil, D.~Momeni and R.~Myrzakulov,
  Eur.\ Phys.\ J.\ C {\bf 72}, 2137 (2012)
  [arXiv:1210.0001 [physics.gen-ph]].
\bibitem{Yang}
R. -J. Yang,
,
Europhys. Lett.
93
, 60001 (2011)
[arXiv:1010.1376 [gr-qc]].
\bibitem{PLB}
K.~Bamba, S.~D.~Odintsov and D.~Sáez-Gómez,
  Phys. Rev. D {bf 88}, 084042 (2013)
  [arXiv:1308.5789 [gr-qc]].
\end{thebibliography}
\end{document}